# Digital imaging of charged particle track structures with a low-pressure optical time projection chamber

U. Titt[§*], V. Dangendorf[§], H. Schuhmacher[§]

[§] *Physikalisch-Technische Bundesanstalt, Bundesallee 100, 38116 Braunschweig, Germany*
[*] *Technische Universität Braunschweig, Hans-Sommer-Straße 10, 38106 Braunschweig, Germany*

We describe first results obtained with a track structure imaging system for measuring the ionisation topology of charged particles in a low-pressure gas. The detection method is based on a time projection chamber (TPC) filled with low-pressure triethylamine (TEA). Images of ionisation tracks of electrons, protons, and heavier ions are presented and analysed.

## 1. Introduction

The understanding of radiation-induced effects in living tissue requires detailed microscopic information about the ionisation topology along the tracks of charged particles with a resolution of the order of several nanometers. In this contribution we describe an imaging system which is capable of measuring these distributions in a low-pressure gas, similar to tissue in its atomic composition. Details of the instrument's application in dosimetry, microdosimetry and neutron spectrometry as well as the measurement principle can be found in [1].

In this paper we present recently introduced detector improvements which allow operation in a self-triggered mode as well as a higher gain and a larger dynamic range to simultaneously measure minimum ionising particles and light ions (p, d, α) at high energy loss close to the Bragg maximum. Furthermore, we show measurement results of two-dimensional images of charged particle tracks like electrons, protons, He and N ions. These measurements enable the determination of the energy losses (d$E$/d$x$ or LET), which are compared to calculated data. Finally we show results of an analysis to identify different particles of similar energy loss by measuring the lateral width of the ionisation track.

## 2. Instrument and basic properties

The method is based on a time projection chamber with a parallel drift field, parallel-plate charge and light amplification steps and optical readout (Optical Avalanche Chamber, OPAC).

Fig. 1 shows the experimental setup. A detailed description of the instrument, its readout facilities and the basic properties can be found in [1]. The chamber is operated with triethylamine (TEA) vapour at a pressure of 10 hPa. At that pressure, the scaling ratio (i.e. the ratio of a length in the gas to the corresponding length in tissue, which is obtained by the ratio of tissue density to gas density) is equal to $2.5 \cdot 10^4$.

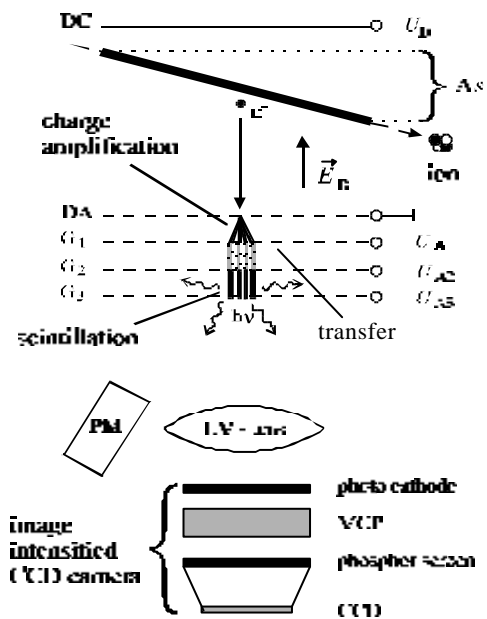

Fig. 1: Schematic view of the optical TPC

The spatial resolution of the chamber is limited by electron diffusion during the drift in the 10 cm wide

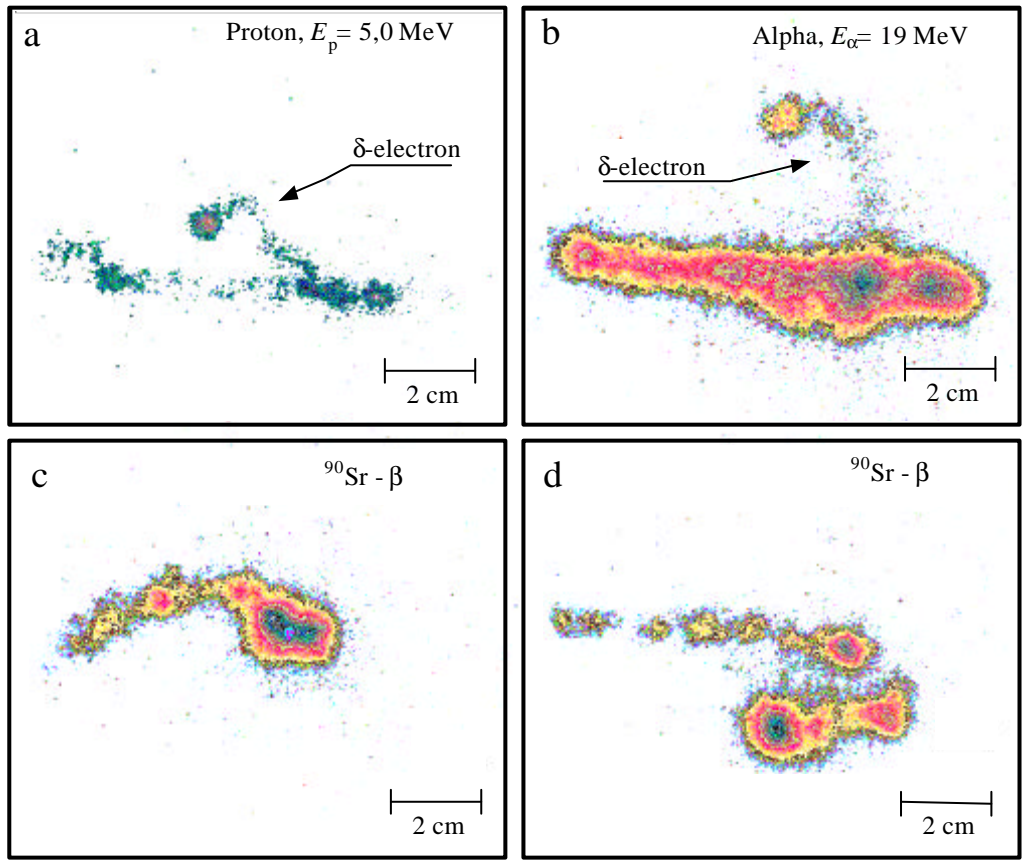

Fig. 2: Samples of measured particle tracks in the OPAC. For the ions (a, b) only such rare events were selected, which show a high energy (several keV) δ-electron emission from the track core. The ion tracks were obtained with a double step amplification [1] while the $^{90}$Sr -β - tracks were measured in the new, self-triggered separated amplification gap device, which provides higher light gain

sensitive volume of the chamber. The transverse diffusion for a 10 cm long drift path in 10 hPa TEA was measured to be in its minimum 3.0 mm, which corresponds to 120 nm in tissue. The longitudinal diffusion amounts to 2.7 mm (i.e. 110 nm in tissue density). Details of the measurements of electron transport in pure low-pressure TEA can be found elsewhere [1].

The device previously described [1] was capable only of measuring in an externally triggered mode. This implies that a separate detector registers the passage of an ionising particle through the OPAC and triggers the optical readout. It was previously planned to use the primary scintillation light which precedes the secondary scintillation by several 10 ns to trigger the system.

The time difference between primary and secondary scintillation would provide the position of the primary charge in the drift direction [2]. However, efficient detection of the primary scintillation light could not be achieved, even in an improved device, where 6 photomultipliers simultaneously view the interaction volume of the chamber.

Therefore, to achieve self-triggered operation of the OPAC, the two amplification stages were separated by an additional 6 mm transfer gap (see fig.1) which introduces a delay of 100–150 ns (depending on the drift voltages) between the rise of the charge signal in the first amplification step and the beginning of the light emission in the secondary scintillation stage. The charge signal of the first stage is then used to trigger the light detection system for the scintillation light produced in the second stage.

Another benefit of the separation of the two stages is the decoupling of the amplification regions. Otherwise photonic and ionic feedback would cause secondary avalanches and eventually an early breakdown in the detector. Thus, a wider stable dynamic range at high gain was accessible. As an example of the high gain achieved with this method, see fig. 2, where the track images of heavily ionising particles (a, b) were obtained with the former doublestep structure at maximum stable gain, while the images of weakly ionising $^{90}$Sr β – particles and its secondaries were made with the separated amplification steps (c, d). The much superior light yield, which can be obtained with the new device is particularly obvious at the end points of the fast electron tracks, which have similar ionisation densities.

## 3. Measurement and results

Fig.3 shows the CCD image (upper part), and the corresponding time projection measurement (lower part) of an 18 MeV nitrogen ion, which crossed the sensitive volume of the detector at an angle of 45° with respect to the optical axis. The energy of the heavy ion significantly decreases in the gas volume of the detector, which leads to an increasing specific energy loss $dE/dx$ during the slow down of the particle. This corresponds to the increasing light yield from the right to the left side in the CCD-image in fig. 3 and, correspondingly, the larger amplitude ($U_A$) of the photomultiplier anode signal for shorter drift times $\Delta t$. The maximum $dE/dx$ is at closest distance from the amplification stage and therefore the first to be displayed. A comparison of the CCD image and the anode pulse trail shows that statistical fluctuations in the ionisation density (clusters), visible as nodes in the CCD image, find corresponding peaks in the anode signal. The duration of the light emission was measured to be 1975 ns (50% of trailing and leading edge). The applied drift field of 6 V cm$^{-1}$ hPa$^{-1}$ leads to an electron drift velocity of 4.7 cm μs$^{-1}$ [1]. With this drift velocity a projected track length parallel to the optical axis ($\Delta s$ in fig.1) of 9.3 cm is calculated.

Fig.4 shows the proportionality of the energy loss $dE/dx$ of various particles (p, α, N, Kr) at low energy, and the light yield $Y_L$ per mm path length. Each data point was obtained by averaging of about 50 tracks. The $dE/dx$ values were calculated with TRIM92 [3]. The heavy ions (N, Kr) were observed close to the Bragg maximum, leading to relatively large uncertainties in the calculated $dE/dx$ due to the large energy loss fluctuation (straggling) of the particle on its way through the foils and gas volumes preceding the sensitive volume of the detector.

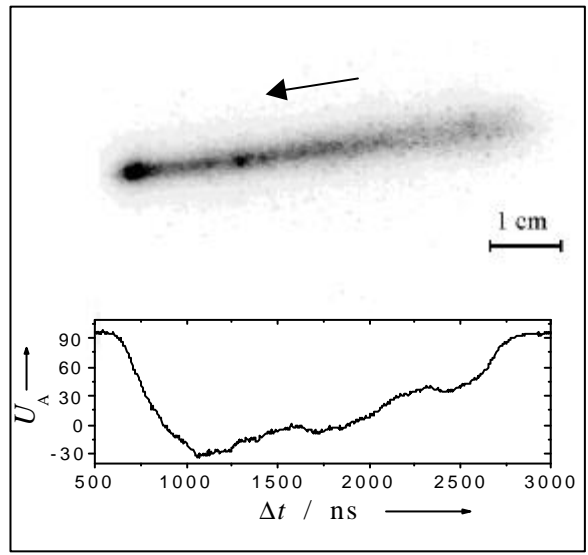

Fig. 3: Track of an 18 MeV N-ion in 10 hPa TEA and the corresponding time projection measurement. The arrow indicates the flight direction of the ion

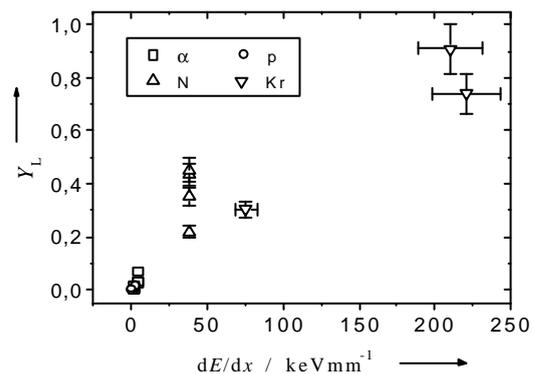

Fig. 4: Measured light yield $Y_L$ (in arbitrary scale) of heavy ion events as a function of the calculated $dE/dx$ from [3]

To identify the particles, the energy loss d$E$/d$x$, and the velocity of the particle have to be measured simultaneously. According to the binary encounter theory of ion-induced δ-electron emission [4], the energy of the δ-electrons varies with the velocity of the ion with

$$E_e = 4\frac{m_e}{m_T} E_T \cos^2 \vartheta_e, \qquad (1)$$

with the electron energy $E_e$, its mass $m_e$, the particle energy $E_T$, its mass $m_T$, and the scattering angle $\vartheta_e$. The maximum energy is of the order of

$$E_e = 4\frac{m_e}{m_T} E_T. \qquad (2)$$

From eq. (2) one expects a correlation between the energy distribution of the ionisation electrons and the particle velocity, namely that heavier particles produce δ-electrons of higher average energy compared to lighter ions of the same d$E$/d$x$. Due to multiple scattering, the flight direction of electrons, which is originally peaked in forward direction, is rapidly randomised. The difference in the δ-electron energy distribution of fast and slow ions of the same d$E$/d$x$ must then lead to a notable difference in the transverse ionisation density distribution. Examples of the transverse ionisation density distribution (expressed as lateral dose distributions) for different particles of the same d$E$/d$x$ can be found in [5].

To directly compare different particles with the same d$E$/d$x$, in particular to identify the particles due to differences in the lateral ionisation density distribution, a series of measurements with deuterons and α-particles was performed. The kinetic energy of the deuterons was 2.13 MeV, 1.03 MeV, 710 keV and 550 keV, the kinetic energy of the corresponding α–particle was 25.6 MeV, 13.9 MeV, 11.0 MeV, 8.25 MeV and 5.44 MeV respectively.

The results of the comparison between the calculated d$E$/d$x$ and the light yield per mm path length $Y_L$ of this experiment are shown in fig.5. Two different amplification parameters were used, resulting in two different slopes of the linear dependence. For each pair of particles of the same d$E$/d$x$, the difference in velocity is analysed by measuring the extension of the transverse ionisation density distribution. This extension is obtained by generating a radial ionisation density histogram from each measured track (see fig. 6) and comparing the overall integral $Y$ of the distribution with the sum $Y_{W1}+Y_{W2}$ of the contents of the wings (see hatched areas in fig. 6). The parameter δ in fig. 6 was chosen to be 25 pixels, corresponding to about 5 mm in the detector. For each kind of particle and each d$E$/d$x$, about 50 tracks were investigated. The results are displayed in fig. 7. The x-axis shows the event number and the y-axis the ratio $R$ as defined in fig.6.

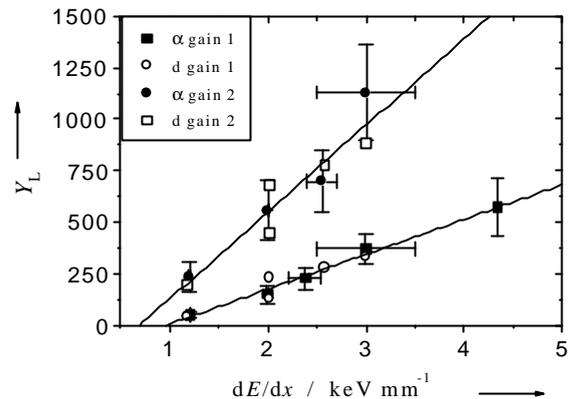

Fig. 5: Linear dependence of light yield per mm track length $Y_L$ (in arbitrary scale) of low energy deuterons and α-particles on the deposited energy per mm path length d$E$/d$x$

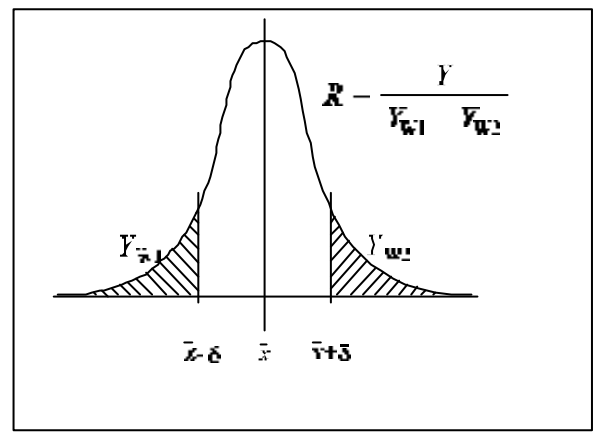

Fig. 6: Transverse ionisation electron distribution of a particle track

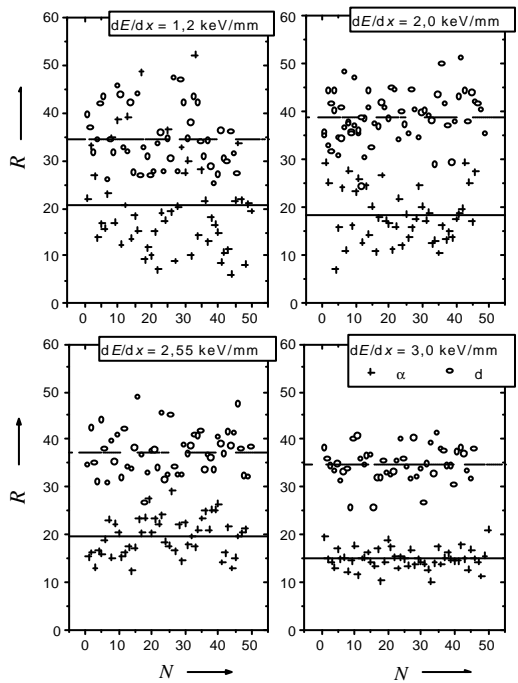

Fig. 7: Comparison of the radial ionisation density, expressed by parameter $R$ (see fig.6) of different particles (d, α) with the same d$E$/d$x$. Each data point represents R of an individual track. Symbols: +: α - particles, o: deuterons

It is clearly visible that for particles of a higher d$E$/d$x$, the ratio $R$ gives an unambiguous measure of the velocity. At lower d$E$/d$x$, the relatively low ionisation electron yield, in combination with clustering effects, lead to an overlap of the measured distributions, and identification of the particles is possible only on the basis of a statistical analysis.

## 4. Conclusion

We have presented a novel technique, based on principles developed for high energy physics experiments [6,7] for application in applied radiation protection (dosimetry and neutron spectrometry) and fundamental research of radiation action in matter and living tissue in particular.

We have shown that for light particles (p, d, α) at low energies particle separation based on d$E$/d$x$ (or LET) analysis and lateral ionisation distribution is feasible. In the further development of the method we will try to derive microdosimetric quantities like proximity functions and lineal energy distributions of virtually defined spheres from the imaged particle tracks and apply this information to derive radiation protection quantities in environments with mixed radiation fields like in nuclear power stations, near high energy accelerators or at high altitude.